\documentclass[12pt]{article}
\usepackage{a4}
\usepackage{amsmath}
\usepackage{bbm,bm,boxedminipage}
\usepackage[latin1]{inputenc}

\def\mathbb#1{\mathbbm{#1}}

\def\+{{+\!\!\!+}}
\def\-{=} 
\def\pp{{\mbox{\tiny${}_{\stackrel\+ =}$}}}

\def\eps{\epsilon}

\def\ph#1{\phantom{#1}}

\def\d{{\rm d}}
\def\del{\delta}
\def\i{{\rm i}}
\def\half{{\frac{1}{2}}}

\def\tsfrac#1#2{{\textstyle \frac{{#1}}{{#2}}}}

\def\+{{+\!\!\!+}}
\def\-{=}

\def\e{\epsilon}

\def\ph#1{\phantom{#1}}

\def\d{{\rm d}}

\ifx\bm\undefined
\else
\fi

\ifx\mathbbm\undefined
  
\else
  
\fi
\ifx\bm\undefined
\else 
\fi

\def\matrix#1#2{\left(\begin{array}{#1}#2\end{array}\right)}

\def\gcgMatrix#1{\matrix{cc}{#1}}
\def\genmatrix#1{\gcgMatrix{#1}}

\ifx\mathbbm\undefined
  
\else
  
\fi

\def\i{{\rm i}}

\def\Q{{\bf Q}}
\def\transl{{\bf P}}
\def\H{{\bf H}}
\def\del{\partial}

\def\GJ{{\cal J}}
\def\GJt{{\tilde{\cal J}}}
\def\GG{{\cal G}}
\def\GJt{{\tilde {\cal J}}}
\def\GK{{\cal K}}
\def\GKt{{\tilde \GK}}

\begin{document}
\def\switchtitlefootnote{\renewcommand{\thefootnote}{\fnsymbol{footnote}}}
\def\switchtextfootnote{\renewcommand{\thefootnote}{\arabic{footnote}}}

\def\abstracttext{We propose the definition of (twisted) generalized
hyperkähler geometry and its relation to supersymmetric non-linear sigma
models. We also construct the corresponding twistor space.  }

%-----------------------------------------------------------
% Title page
%-----------------------------------------------------------
\switchtitlefootnote
\pagestyle{empty}
\begin{titlepage}

\begin{center}

\vspace*{-1.0cm}
{\hbox to \hsize{\hfill UUITP-12/06}}
{\hbox to \hsize{\hfill ~}}
\vspace*{1.0cm}

\vspace*{1.5cm}
{\Large \bf 
        Generalized Hyperkähler Geometry\\
	and Supersymmetry}

\vspace*{1.5cm}
{       Andreas Bredthauer\footnote[1]{\parbox[t]{12cm}{\tt 
        Andreas.Bredthauer@teorfys.uu.se}}}\\[1cm]

{\em	
        Department for Theoretical Physics\\
        Uppsala University\\
        Box 803, SE-751\,08 Uppsala, Sweden}

\vspace*{1cm}
{\small% August 15, 2006
}

\vspace*{5cm}
{\small \bf Abstract}\\[0.5cm]

\parbox{12.5cm}
{\small
%\hspace*{0.7cm}
\abstracttext
}
\end{center}
\end{titlepage}

\newpage

\setcounter{page}{1}
\switchtextfootnote
\pagestyle{plain}

\section{Introduction}

It is well-known that supersymmetry has a deep relation to geometry
\cite{Zumino:1979et,Alvarez-Gaume:1980vs}. In the context of sigma models the geometry of the target
space is restricted by the amount of supersymmetry on the worldsheet and its
dimension. For example, for a two-dimensional worldsheet with manifest $N=(1,1)$
supersymmetry the sigma model has $N=(2,2)$ supersymmetry if the target manifold
is bi-hermitian \cite{Gates:1984nk}. For around twenty years now the different
possible geometries have been studied
\cite{Howe:1985pm,Buscher:1987uw,Hitchin:1986ea,
Howe:1988cj,Howe:1996kj,Ivanov:2001dn}. For an
overview over the different possibilities, see also \cite{Lindstrom:2004iw}.
Although this classification is known, it was only recently that an elegant
mathematical framework to deal with these geometries was developed. Hitchin
introduced the notion of generalized complex geometry in the context of
generalized Calabi-Yau manifolds with fluxes \cite{Hitchin:2004ut}. Later, it
was clarified by Gualtieri \cite{Gualtieri:2003dx}.  The main idea is to replace
the tangent bundle $TM$ of a manifold $M$ with the sum of the tangent and the
cotangent bundle $TM\oplus T^*M$. This puts metrics and two-forms of the sigma
model on an equal footing and unifies the concepts of complex and symplectic
geometry. It has been shown that bi-hermitian geometry corresponds to a subset
of generalized complex geometry called generalized Kähler geometry. The obvious
question is how generalized complex geometries arise in the sigma model context.
A lot of work has been done in this direction \cite{Lindstrom:2004eh,
Bergamin:2004sk,
Zucchini:2004ta,
Lindstrom:2004iw,
Bredthauer:2005zx,
Zabzine:2005qf,
Calvo:2005ww}
and by now for the generalized Kähler geometry, the picture is rather clear
\cite{Lindstrom:2006ee,Zabzine:2006uz}. In the Lagrangian formulation of the
sigma model generalized Kähler geometry is completely specified by the
definition of a Lagrangian in a manifest $N=(2,2)$ supersymmetric formulation.
The Lagrangian is the generalized Kähler potential \cite{Lindstrom:2005zr}. On
the other hand it has been shown that it can be physically derived as the
geometry admitting $N=(2,2)$ supersymmetry in the phase space formulation of the
sigma model \cite{Bredthauer:2006hf}. In this letter, we elaborate further in
this second direction and derive the conditions for $N=4$ and $N=(4,4)$
supersymmetry. We show how generalized hypercomplex and generalized
hyperkähler geometry arise in the phase space formulation and construct the
twistor space of generalized complex structures that is associated with the
$N=(4,4)$ supersymmetry. 

This letter is organized as follows: In section 2, we review extended
supersymmetry of the $N=(1,1)$ supersymmetric sigma model and the relevant
target space geometries. In section 3, we give an overview on the results for
the phase space formulation of the $N=(1,1)$ sigma model. In section 4 we show
generalized hypercomplex geometry arises in this notion and in section 5 we
combine these results with those of a previous paper \cite{Bredthauer:2006hf}
and give the derive generalized hyperkähler geometry from $N=(4,4)$
supersymmetry. In section 6 we comment shortly on how this result relates to the
Lagrangian formulation of the sigma model. In sections 7 and 8 we define 
the twistor space for the generalized hyperkähler geometry before concluding with a
short discussion in section 9.

\section{$N=(1,1)$ sigma model and extended supersymmetry}
In this section, we review extended supersymmetry of the $N=(1,1)$
supersymmetric sigma model and introduce the necessary notation.
The action for the $N=(1,1)$ supersymmetric sigma model is given by
\begin{align}
  S = \frac{1}{2}\int \d^2\sigma \d\theta^+ \d\theta^- D_+\Phi^\mu D_-\Phi^\nu
      \big( G_{\mu\nu}(\Phi) + B_{\mu\nu}(\Phi) \big). \label{action}
\end{align}
Here, $D_\pm$ are spinorial derivatives with
\begin{align}
  D_\pm = \partial_{\theta^\pm} + \i \theta^\pm \partial_\pp,
&&  D_\pm^2 = \i\partial_\pp, && && \{D_+,D_-\} = 0.
\end{align}
The action is invariant under  a manifest supersymmetry that is given by
\begin{align}
  \delta_0(\eps)\Phi^\mu = -\i(\eps^+Q_+ + \eps^-Q_-)\Phi^\mu, \label{delta0}
\end{align}
where $Q_\pm = \i D_\pm + 2\theta^\pm \partial_\pp$ are the supersymmetry
charges. The action admits an extension to $N=(2,2)$ supersymmetry if the target
space geometry is bi-hermitian \cite{Gates:1984nk}. The additional supersymmetry
transformation is of the form
\begin{align}
  \delta_1(\e)\Phi^\mu = \e^+D_+\Phi^\nu J^\mu_{+\nu} + \e^-D_-\Phi^\nu
  J^\mu_{-\nu}, \label{delta1}
\end{align}
where $J_{\pm}$ are complex structures that satisfy
\begin{align}
  J_{\pm\mu}^\rho G_{\rho\sigma} J_{\pm\nu}^\sigma = G_{\mu\nu}, &&
  \nabla^{(\pm)}_\rho J_{\pm \nu}^\mu = 0. \label{bi-herm}
\end{align}
The connections for the covariant derivatives are given by
\begin{align}
  \Gamma^{(\pm)\mu}_{\nu\rho} = \Gamma^{(0)\mu}_{\nu\rho} \pm %\half
  G^{\mu\sigma}H_{\sigma\nu\rho},
\end{align}
where $\Gamma^{(0)}$ is the Levi-Civita connection for the metric $G_{\mu\nu}$
and $H=\d B$ is the torsion three-form. Explicitly, it is given by
\begin{align}
  H_{\mu\nu\rho} = \half\big( B_{\mu\nu,\rho} + B_{\nu\rho,\mu} +
  B_{\rho\mu,\nu}\big).
\end{align}
Indices separated by a comma define derivatives with respect to the
corresponding spacetime direction $B_{\mu\nu,\rho} = \partial_\rho B_{\mu\nu}$.
We denote the Kähler forms as $\omega_{\pm\mu\nu} =
G_{\mu\rho}J^{\rho}_{\pm\nu}$. In general, the non-manifest
supersymmetry transformation only closes on-shell 
\begin{align}
  [\delta_1(\eps),\delta_1(\tilde\eps)]\Phi^\mu = 2\eps^+\tilde\eps^+\partial_\+
\Phi^\mu + 2\eps^-\tilde\eps^-\partial_\- \Phi^\mu. 
\end{align}

The restrictions on the target manifold geometry increases if we consider
$N=(4,4)$ supersymmetry. In that case, there are three additional
supersymmetries of the form \eqref{delta1}
\begin{align}
  \delta_i(\eps)\Phi^\mu = \eps^+D_+\Phi^\nu J^\mu_{+i\nu} +
     \eps^-D_-\Phi^\nu J^\mu_{-i\nu},~~i=1,2,3.
  \label{deltaPhi-oneJ}
\end{align}
As an implication of the previous discussion, these are (on-shell) supersymmetry
transformations if the six tensors $J_{\pm i}$ are complex structures that
satisfy \eqref{bi-herm}. This is a consequence of the fact that each of the
additional supersymmetry transformations commute with the manifest supersymmetry
\eqref{delta0}. In addition, the transformations have to commute
among themselves in order to satisfy the supersymmetry algebra. Altogether, we
have
\begin{align}
  [\delta_i(\eps),\delta_j(\tilde\eps)] \Phi^\mu =
2\delta_{ij}\big(\eps^+\tilde\eps^+ \partial_\+\Phi^\mu +
\eps^-\tilde\eps^-\partial_\-\Phi^\mu\big),~~i,j=0,1,2,3.
\end{align}
We do not discuss the possibility of central charges here. These relations are
satisfied on-shell if the left- and the right-going complex structures
anticommute among themselves
\begin{align}
  \{ J_{+ i}, J_{+ j} \} = \{ J_{-i}, J_{- j} \} = -2\delta_{ij}1.
\end{align}
They form two hypercomplex structures. The target space geometry is
bi-hypercomplex \cite{Gates:1984nk}. We collect the requirements needed for
$N=(4,4)$ supersymmetry of the action \eqref{action}:
\begin{align}
  J_{\pm i \mu}^\rho G_{\rho\sigma} J_{\pm i \nu}^\sigma = G_{\mu\nu}, &&
  \nabla^{(\pm)}_\rho J^\mu_{\pm\nu} = 0, &&  
  J_{\pm i \rho}^\mu J_{\pm j \nu}^\rho + J_{\pm i \rho}^\mu J_{\pm j \nu}^\rho
  = -2\delta_{ij}\delta^\mu_\nu.
\end{align}

One way to achieve off-shell supersymmetry is if the left- and right-going
complex structures in \eqref{delta1} or \eqref{deltaPhi-oneJ} commute. A
particular case is when the two complex structures are equal and the extended
supersymmetry transformations are of the form
\begin{align}
  \delta_i(\eps)\Phi^\mu = (\eps^+D_+\Phi^\nu + \eps^-D_-\Phi^\nu)J^\mu_{i \nu}.
\end{align}
The conditions for supersymmetry imply that this is only possible if the 
torsion $H$ is zero. In the case of $N=(2,2)$ supersymmetry, the target space 
geometry is Kähler, while the action \eqref{action} admits $N=(4,4)$
supersymmetry on a hyperkähler manifold.

\section{$N=2$ extended supersymmetry in phase space}
In \cite{Zabzine:2005qf} $N=1$ supersymmetric phase space was introduced as the
cotangent bundle $\Pi T^*{\cal L}M$ (with parity reversed) of the superloop
space ${\cal L}M = \{\phi: S^{1,1}\rightarrow M\}$. Here, $S^{1,1}$ is a
supercircle, a circle with an additional Grassmann direction $\theta$.
$\phi^\mu$ is a superfield that embeds the supercircle into the manifold. The
conjugate momenta $S_\mu$ are spinorial fields. Therefore, the cotangent bundle
has reversed parity on its fibers. The superfields have the following expansions
in the odd coordinate
\begin{align}
  \phi^\mu(\sigma,\theta) = X^\mu(\sigma) + \theta \lambda^\mu(\sigma), &&
  S_\mu(\sigma,\theta) = \psi_\mu(\sigma) + \i\theta p_\mu(\sigma),
\end{align}
where $p_\mu$ is the momentum conjugate to $X^\mu$. We use the notation of
\cite{Bredthauer:2006hf}.  The phase space is equipped with a symplectic
structure
\begin{align}
  \omega = \i\int\d\sigma\d\theta \delta S_\mu\wedge \delta\phi^\mu,
\end{align}
where $\delta$ is the de Rham differential on the manifold. The convention for
$\omega$ is chosen in such a way that its purely bosonic part is equal to the
usual definition of the canonical symplectic structure in bosonic phase space
\begin{align}
  \omega|_{\rm bos} = \int\d\sigma \delta X^\mu \wedge \delta p_{\mu}.
\end{align}
The symplectic structure yields a superPoisson bracket
\begin{align}
  \{F,G\} = \i\int\d\sigma\d\theta F \Big( 
\frac{\overleftarrow\delta}{\delta S_\mu}\frac{\overrightarrow\delta}{\delta\phi^\mu}
-\frac{\overleftarrow\delta}{\delta \phi^\mu}\frac{\overrightarrow\delta}{\delta S_\mu}
  \Big) G.
\end{align}
$\omega$ can be twisted by a three-form $H$
\begin{align}
  \omega_H = \i\int\d\sigma\d\theta \Big(\delta S_\mu \wedge \delta \phi^\mu -
  \half H_{\mu\nu\rho}D\phi^\mu \delta\phi^\nu\wedge \delta \phi^\rho\Big).
\end{align}
Here, $\partial$ is the derivative with respect to $\sigma$.
Consequently, this twists the Poisson bracket by $H$. To keep things simple, we
here work with the case $H=0$ and only comment on the changes for $H\neq 0$. The
calculations in that case work out exactly in the same way.

The phase space has two natural operations
\begin{align}
  D = \partial_\theta +\i \theta\partial, &&
  Q = \partial_\theta -\i \theta\partial.
\end{align}
They satisfy the algebra
\begin{align}
  D^2 = \i\partial, && Q^2 = -\i\partial, && \{D,Q\} = 0.
\end{align}
With this, the generator for manifest supersymmetry is given by
\begin{align}
  {\Q}_0(\e) = -\int \d\sigma \d\theta \e S_\mu Q \phi^\mu, \label{Q0}
\end{align}
where $\e$ is an odd parameter. It acts on the fields through the Poisson
bracket
\begin{align}
  \delta(\eps)\phi^\mu = \{\phi^\mu, {\Q}_0(\eps)\} = -\i\eps Q\phi^\mu, &&
  \delta(\eps)S_\mu = \{S_\mu, {\Q}_0(\eps)\} = -\i QS_\mu. \label{Q-delta}
\end{align}
Being a supersymmetry generator, it satisfies the supersymmetry algebra
\begin{align}
  \{\Q_0(\e),\Q_0(\tilde\e)\} = \transl (2\e\tilde \e), &&
  {\transl}(a) = \int \d\sigma \d\theta a S_\mu \del \phi^\mu, \label{PB}
\end{align}
where $\transl(a)$ is the generator of $\sigma$-translations. Any additional
supersymmetry that is generated by some $\Q_1(\e)$ has to satisfy the brackets
\begin{align}
  \{\Q_0(\e), \Q_1(\tilde\e) \} = 0, &&
  \{\Q_1(\e), \Q_1(\tilde\e) \} = \transl(2\e\tilde\e). \label{Q1-PB}
\end{align}
The condition for which these conditions are satisfied was found in \cite{Zabzine:2005qf}. The
form of $\Q_1(\eps)$ is determined by dimensional arguments
\begin{align}
  \Q_1(\e) = -\frac{1}{2}\int \d\sigma\d\theta\e\Big(2D\phi^\mu S_\nu J^\nu_\mu
  +D\phi^\mu D\phi^\nu L_{\mu\nu} + S_\mu S_\nu P^{\mu\nu}\Big), \label{Q1}
\end{align}
where the tensors can be conveniently combined into a map 
${\cal J}: T\oplus T^* \rightarrow T\oplus T^*$ given by
\begin{align}
  {\cal J} = \left(\begin{array}{cc}-J&\ph{0}P\\ \ph{0}L&\ph{0}J^t
   \end{array}\right).\label{GJ}
\end{align}
$\Q_1(\eps)$ is the generator of a supersymmetry transformation if the target
space is generalized complex and ${\cal J}$ is a generalized complex structure.

For $H\neq 0$, we can construct the generators of supersymmetry in a similar
way. With $H=\d B$, this can be achieved by replacing
\begin{align}
  S_\mu \rightarrow S_\mu - B_{\mu\nu}D\phi^\nu
\end{align}
in the definitions of the supersymmetry generators, for example
\begin{align}
  \Q_0 = -\int\d\sigma\d\theta\eps(S_\mu - B_{\mu\nu}D\phi^\nu)Q\phi^\mu.
\end{align}
Since we are forced to use the twisted version of the Poisson bracket as well,
the transformation on the fields \eqref{Q-delta} remains unchanged.  Concerning
the additional supersymmetry, we may afterwards rename the tensors in such a way
that $\Q_1$ remains in the form \eqref{Q1}. ${\cal J}$ is then a twisted
generalized complex structure. 

\section{$N=4$ extended supersymmetry in phase space}
In the previous section we reviewed the steps that lead to the condition that
$N=2$ extended supersymmetry in phase space is possible if the target space is
a generalized complex manifold. The generator for the extended supersymmetry is
given in \eqref{Q1}. Here, we discuss the necessary conditions for to have two
such extended supersymmetries with generators $\Q_1(\e)$ and $\Q_2(\e)$ of the
form \eqref{Q1}. 

We show that the target space geometry has to be generalized hypercomplex. We
define generalized hypercomplex geometry by three generalized complex structures
that satisfy the algebra of quaternions
\begin{align}
  {\cal J}_3 = {\cal J}_1 {\cal J}_2.
\end{align}

Following the discussion of the previous section, $\Q_1(\e)$ and $\Q_2(\e)$ are
generators of supersymmetry if we can relate them to two generalized complex
structures ${\cal J}_1$ and ${\cal J}_2$. In addition to \eqref{PB} and
\eqref{Q1-PB}, the Poisson bracket of $\Q_1(\e)$ with $\Q_2(\e)$ has to vanish
\begin{align}
  \{\Q_1(\e),\Q_2(\tilde\e)\} = 0. \label{Q1Q2}
\end{align}
We can collect all the Poisson brackets in the following way:
\begin{align}
  \{\Q_i(\e),\Q_j(\tilde\e)\} = \delta_{ij}\transl(2\eps\tilde\eps),~~i=0,1,2.
\end{align}
The calculation of the bracket \eqref{Q1Q2} is tedious but results in the two
conditions for the generalized complex structures
\begin{align}
  \{ {\cal J}_1,{\cal J}_2\} = 0, && {\cal N}({\cal J}_1,{\cal J}_2) = 0.
\label{ghcs}
\end{align}
Here, ${\cal N}({\cal J}_1,{\cal J}_2)$ is the (generalized) Nijenhuis concomitant 
of $\GJ_1$ and $\GJ_2$. It is given by
\begin{multline}
  {\cal N}(\GJ_1,\GJ_2) = [\GJ_1(u+\xi),\GJ_2(v+\eta)] -
\GJ_1[u+\xi,\GJ_2(v+\eta)] \cr -
\GJ_2[\GJ_1(u+\xi),v+\eta] +  \GJ_1\GJ_2[u+\xi,v+\eta] - (\GJ_1\leftrightarrow
\GJ_2),
\end{multline}
where the bracket is the Courant bracket and $u+\xi, v+\eta$ are sections of
$TM\oplus T^*M$. 

It follows that $\GJ_3=\GJ_1\GJ_2$ is the third generalized complex
structure. Its integrability is guaranteed by the vanishing of the
(generalized) Nijenhuis concomitant and integrability of $\GJ_1$ and $\GJ_2$.
This proves that the target manifold is generalized hypercomplex.

The existence of $\GJ_3$ implies the existence of a third additional
supersymmetry with generator $\Q_3(\eps)$ of the form \eqref{Q1} and we conclude
that a generalized hypercomplex manifold extends the supersymmetry of the sigma
model phase space to $N=4$.

For the case $H\neq 0$ the calculations remain true but we have to replace the
Poisson bracket by its twisted version. The target space geometry is then
twisted generalized hypercomplex.

\section{Generalized Hyperkähler Geometry}
The previous discussion was completely model independent. In this section, we
combine the results with those of a previous paper \cite{Bredthauer:2006hf}.
There, it was shown that from the sigma model point of view, the relation
between generalized Kähler and bi-hermitian geometry follows from the
equivalence of the Hamilton and the Lagrange description of the sigma model. 
We start with a short recapitulation of those results. The sigma model
Hamiltonian for \eqref{action} is obtained by performing one of the
$\d\theta$-integrations\footnote{See also \cite{Malikov:2006rm} for a
mathematically more rigid derivation.}. To this extend, we introduce new
Grassmann coordinates $\theta^0$ and $\theta^1$ such
that
\begin{align}
  \theta^{0,1} = \frac{1}{\sqrt{2}}(\theta^+\mp\i\theta^-), &&
  D_{0,1} = \frac{1}{\sqrt{2}}(D_+\pm\i D_-), &&
  Q_{0,1} = \frac{1}{\sqrt{2}}(Q_+\pm\i Q_-).
\end{align}
We define the $N=1$ component fields of $\Phi^\mu$ by
\begin{align}
  \phi^\mu = \Phi^\mu|_{\theta^0 = 0}, &&
  S_\mu = G_{\mu\nu}D_0\Phi^\mu|_{\theta^0 = 0}
\end{align}
and denote $G_{\mu\nu}(\phi) = G_{\mu\nu}(\Phi)|$, $D=D_1|$ and
$\partial=\partial_\sigma$. Performing the $\d\theta^0$ integral in the usual
way by replacing $\int\d\theta^0\rightarrow D_0$ and taking the $\theta^0=0$
component we obtain the action \eqref{action} in terms of the $N=1$ superfields
\begin{align}
  S = \int \d^2\sigma \d\theta \Big(\i S_\mu\partial_0\phi^\mu - \frac{1}{2}\big(
    \i\del\phi^\mu D\phi^\nu + S_\mu DS_\nu G^{\mu\nu} + S_\mu D\phi^\nu S_\rho
G^{\sigma\rho}\Gamma^\mu{}_{\nu\sigma}\big)\Big).
\end{align}
Here, we focus on the case $B_{\mu\nu}=0$. The action has the typical form of a
Legendre transformation where the first term says that $S_\mu$ is
the conjugate momentum for $\phi^\mu$ and the second term yields the Hamiltonian
\begin{align}
  {\H} = \frac{1}{2}\int\d\sigma\d\theta \big(
    \i\del\phi^\mu D\phi^\nu + S_\mu DS_\nu G^{\mu\nu} + S_\mu D\phi^\nu S_\rho
G^{\sigma\rho}\Gamma^\mu{}_{\nu\sigma}\big).
\end{align}
${\H}$ is invariant under the manifest supersymmetry transformation
\eqref{Q0} with $Q\equiv Q_1|$. The second manifest supersymmetry of the original action
\eqref{action} is non-manifest in this formulation. It is given by
\begin{align}
  \tilde \delta_0(\eps)\phi^\mu = \eps G^{\mu\nu}S_\nu, &&
  \tilde \delta_0(\eps)S_\mu = \i\eps G_{\mu\nu}\del \phi^\nu + \eps S_\nu
  S_\rho G^{\nu\sigma}\Gamma^{\rho}{}_{\mu\sigma}.
\end{align}
In \cite{Bredthauer:2006hf} it is shown that ${\H}$ is invariant under the
additional supersymmetry if the target space geometry is generalized Kähler:
\begin{align}
  \{\Q_1(\eps),{\H}\} = 0 \label{Q1H}
\end{align}
implies that $\GJ$ commutes with the generalized metric
\begin{align}
  \GG = \genmatrix{ & G^{-1} \\ G & }. \label{genG}
\end{align}
As a consequence, $\GJt = \GG\GJ$ is an additional generalized complex structure
with supersymmetry generator $\tilde \Q_1(\eps)$ of the form \eqref{Q1} such
that
\begin{align}
  [\GJ,\GJt] = 0, && \{\Q_1(\eps),\tilde\Q_1(\tilde\eps)\} = 2\i\eps\tilde\eps
  {\H}.
\end{align}

We now show that the Hamiltonian has $N=(4,4)$ supersymmetry if the geometry is
generalized hyperkähler. We call a manifold generalized hyperkähler if it admits
six generalized complex structures $\GJ_i, \GJt_i$ and a generalized metric
$\GG$ that satisfy the
algebra of bi-quaternions $Cl_{2,1}(\mathbb{R})$
\begin{align}
  &\GJ_i \GJ_j = -\delta_{ij} 1_{2d} + \e_{ijk} \GJ_k &
  &\GJt_i \GJt_j = -\delta_{ij} 1_{2d} + \e_{ijk} \GJ_k \cr
  &\GJ_i \GJt_j = -\delta_{ij} \GG + \e_{ijk} \GJt_k&
  &\GJt_i \GJ_j = -\delta_{ij} \GG + \e_{ijk} \GJt_k.  \label{ghks}
\end{align}
This definition coincides with \cite{Huybrechts:2003ak,Goto:2005}. However, we
derive this definition from the sigma model.

The Hamiltonian is invariant under the three additional supersymmetries of the
previous sections if it satisfies
\begin{align}
  \{\Q_i(\e), {\H}\} = 0, ~~ i=1,2,3.
\end{align}
As in \eqref{Q1H}, this is the case if the generalized complex structures $\GJ_i$
commute with $\GG$
\begin{align}
  [\GJ_i,\GG] = 0.
\end{align}
According to the above, this induces three generalized complex structures
$\GJt_i =\GG\GJ_i$. Each of the triples $\{\GJ_i,\GG,\GJt_i\}$ for
$i=1,2,3$ form a generalized Kähler structure. The three generalized complex
structures $\GJt_i$ are associated to three additional supersymmetry generators
$\tilde\Q_i$. The generators satisfy the algebra
\begin{align}
  \{\Q_i(\eps),\tilde\Q_j(\tilde\eps\} = 2\i\delta_{ij}\eps\tilde\eps{\H}, &&
  \{\Q_i(\eps),\Q_j(\tilde\eps)\} =\delta_{ij}\transl(2\eps\tilde\eps).
\end{align}
A straightforward calculation shows that these brackets are equivalent to the 
relations \eqref{ghks} and integrability of $\GJ_i$ and $\GJt_i$.  We conclude
that the sigma model Hamiltonian admits $N=(4,4)$ supersymmetry if the target
space is generalized hyperkähler.

For $H\neq 0$, the Hamiltonian is given by
\begin{multline}
  {\H} = \frac{1}{2}\int\d\sigma\d\theta \Big(
    \i\del\phi^\mu D\phi^\nu + S_\mu DS_\nu G^{\mu\nu} + S_\mu D\phi^\nu S_\rho
G^{\sigma\rho}\Gamma^\mu{}_{\nu\sigma} \cr
  - \frac{1}{3}H^{\mu\nu\rho}S_\mu S_\nu S_\rho + D\phi^\mu D\phi^\nu S_\rho
H_{\mu\nu}{}^\rho \Big)
\end{multline}
and admits $N=(4,4)$ supersymmetry for a twisted generalized hyperkähler
geometry which is defined in analogy to generalized hyperkähler geometry but
with twisted generalized complex structures.

\section{Relation to the Lagrangian formulation}
The three different generalized Kähler structures $\{\GJ_i,\GG,\GJt_i\}$
correspond to bi-hermitian geometries, where the metric is read off from
\eqref{genG} and the complex structures are given via the relation
\cite{Gualtieri:2003dx}
\begin{align}
  \GJ_i = \frac{1}{2}\genmatrix{-(J_{+i}+J_{-i}) & -(\omega_{+i}^{-1} -
   \omega_{-i}^{-1}) \\ \omega_{+i} - \omega_{-i} & (J_{+i}+J_{-i})^t} \cr
  \GJt_i = \frac{1}{2}\genmatrix{-(J_{+i}-J_{-i}) & -(\omega_{+i}^{-1} +
   \omega_{-i}^{-1}) \\ \omega_{+i} + \omega_{-i} & (J_{+i}-J_{-i})^t},
\label{GKG=BH}
\end{align}
where $\omega_{\pm i}=GJ_{\pm i}$ are the Kähler forms. From the bi-quaternion
algebra it is easy to see that $J_{+i}$ and $J_{-i}$ form two independent
hypercomplex structures with
\begin{align}
  \{J_{+1}, J_{+2}\} = 0, && \{J_{-1},J_{-2}\} = 0
\end{align}
but with nothing implied for the commutation relations of $J_{+i}$ and $J_{-j}$.
For the case $H\neq 0$ we obtain in addition the relations of section 2. We
conclude that (twisted) generalized hyperkähler geometry is the phase space
equivalent to bi-hypercomplex geometry.

\section{Twistor Space of Generalized Complex Structures}
In this section, we define the twistor space of generalized complex structures
that is associated to the $N=(4,4)$ supersymmetry of the sigma model
Hamiltonian. The idea of a twistor space is to encode the geometric properties
of the target manifold $M$ in the holomorphic structure of a larger manifold,
the twistor space. The original idea goes back to Penrose \cite{Penrose:1976}
and Salamon \cite{Salamon:1982,Salamon:1986}. We here follow the same approach
as in the definition of the twistor space for hyperkähler geometry \cite{Hitchin:1986ea}. 
Twistor spaces of generalized complex structures and generalized Kähler
structure are also discussed in \cite{Davidov:2005, Davidov:2006} in order to
find examples of generalized complex and generalized Kähler structures that are
not induced by complex, symplectic and Kähler structures.  Before discussing the
twistor space for the generalized hyperkähler geometry, we first review the
results for hyperkähler geometry.  Given a hypercomplex structure $J_1$, $J_2$,
$J_3$ the linear combination
\begin{align}
  K = c^1 J_1 + c^2 J_2 + c^3 J_3
\end{align}
is a complex structure if $\vec{c}$ lies on the unit sphere: $c^2 = 1$. This
sphere can be identified with $\mathbb{C}P^1$. $\mathbb{C}P^1$ is usually
represented as $\mathbb{C}^2$ with coordinates $(\zeta,\tilde\zeta)$ and the
identification $(\zeta,\tilde\zeta)\simeq(\lambda\zeta,\lambda\tilde\zeta)$ for
$\lambda\neq 0$. Therefore, we can cover it with two sheets of coordinates
$(\zeta, 1)$ and $(1,\tilde \zeta)$ such that $\tilde \zeta = \zeta^{-1}$ in the
overlapping region. In these coordinates,
\begin{align}
 K = \frac{1-\zeta\bar\zeta}{1+\zeta\bar\zeta}J_1 +
    \frac{\zeta+\bar\zeta}{1+\zeta\bar\zeta}J_2 +
     \i\frac{\zeta-\bar\zeta}{1+\zeta\bar\zeta}J_3.
\end{align}
The twistor space of complex structures is the product space $M\times S^2$, such
that at any point $p\in M$, $S^2$ parametrized the space of complex structures
on $T_p M$. A complex structure for the whole manifold is then given by the pair
\begin{align}
  {\mathbf{K}}_\zeta = \left(\frac{1-\zeta\bar\zeta}{1+\zeta\bar\zeta}J_1 +
    \frac{\zeta+\bar\zeta}{1+\zeta\bar\zeta}J_2 +
    \i\frac{\zeta-\bar\zeta}{1+\zeta\bar\zeta}J_3, I\right),
\end{align}
where $I$ is the is the standard complex structure on the sphere. This construction
allows to define hyperkähler geometry in terms of an abstract parameter space.

We now define the twistor space of generalized complex structures in a
completely analogous way. Given the six generalized complex structures $\GJ_i$ and
$\GJt_i$ of the previous section, we find that the linear combinations that 
define generalized complex structures are given by the relation
\begin{align}
  \GK = \tsfrac{1}{2}(c^i+d^i)\GJ_i + \tsfrac{1}{2}(c^i-d^i)\GJt_i, &&
  \vec{c}^2 = \vec{d}^2 = 1. \label{GK=}
\end{align}
The space of generalized complex structures for a generalized hyperkähler
structure is parametrized by $S^2\times S^2$. In $\mathbb{C}P^1\times
\mathbb{C}P^1$ coordinates $z,w$, the vectors $\vec{c}$ and $\vec{d}$ are given
by
\begin{align}
  \vec{c}=\left(\frac{1-z\bar z}{1+z\bar z}, \frac{z+\bar z}{1+z\bar z},
   \frac{\i(z-\bar z)}{1+z\bar z}\right),&&
  \vec{d}= \left(\frac{1-w\bar w}{1+w\bar w}, \frac{w+\bar w}{1+w\bar w},
   \frac{\i(w-\bar w)}{1+w\bar w}\right).
\end{align}
Since the generalized complex structures $\GJ_i$, $\GJt_i$ are a realization of
the bi-quaternionic algebra, it follows that $\GK^2=-1$ and $\tilde\GK = \GG\GK$
where $\GG$ is the generalized metric. The generalized metric $\GG$ acts on the 
parameter space by letting $\vec{d}\rightarrow -\vec{d}$. In the $\mathbb{C}P^1$
coordinate $w$, this corresponds to the anti-podal map
\begin{align}
\tau_w : w \rightarrow -\tilde w^{-1}
\end{align}
that changes the orientation of the $w$-sphere. The ordinary complex structures
for the two spheres $I_z$ and $I_w$ define a complex structure $J_S$ for
$S^2\times S^2$. This
complex structure induces a generalized complex structure on $T(S^2\times
S^2)\oplus T^*(S^2\times S^2)$ by
\begin{align}
  {\cal J}_S = \left( \begin{array}{cc}-J_S&0 \\ 0& J_S^t \end{array} \right).
\end{align}
A generalized complex structure for the combined space $M\times S^2\times S^2$
is then given by 
\begin{align}
  {\bf J} = (\GK(z,w), {\cal J}_S). \label{bfJ}
\end{align}
The proof that ${\bf J}$ is integrable is presented in the next section using
the formulation of generalized complex structures in terms of pure spinor
lines. It is an interesting question, if ${\cal I}$ can be chosen in a more
general way in this context. Generalized complex structures for $S^2\times S^2$
were explicitly defined in \cite{Hitchin:2005cv}.
 
The triples $\{\GK, \GG, \GKt = \GG\GK\}$ form different generalized Kähler structures.
The two spheres parametrize the space of ordinary left- and right-complex
structures on $TM$. We can clarify this by introducing 
\begin{align}
  \GJ_i^{(\pm)} = \frac{1}{2}\big(\GJ_i\pm\GJt_i\big) = \frac{1}{2}\big(1 \pm
  \GG\big) \GJ_i.
\end{align}
These are the projections of the generalized complex structures on the $\pm$
eigenspaces of $\GG$. Explicitly and with relation \eqref{GKG=BH}, they are given by
\begin{align}
  \GJ_i^{(\pm)} = \frac{1}{2}\left(\begin{array}{cc}
    -J_{\pm i}& -\omega^{-1}_{\pm i} \\
    \omega_{\pm i}& J_{\pm i}^t
  \end{array}\right).
\end{align}
With this, \eqref{GK=} becomes
\begin{align}
  \GK = c^i\GJ_i^{(+)} + d^i\GJ_i^{(-)}.
\end{align}
We indeed find that $\vec{c}$ and $\vec{d}$ parametrize the two sets of complex
structures $J_{+i}$ and $J_{-i}$.

\section{Pure spinors}
It remains to show that ${\bf J}$ as defined in \eqref{bfJ} is indeed a
generalized complex structure. In order to see this, we reformulate the previous
discussion in the pure spinor language. Since $TM\oplus T^*M$ always admits a
$Spin(d,d)$ structure that is isomorphic to the exterior algebra $\wedge T^*M$,
we can associate the $+\i$ eigenspace $L$ of a generalized complex structure
$\GJ_1$ with the annihilation space of a spinor $\varphi$ such that for the sections
$u+\xi$ of $L$,
\begin{align}
  (u+\xi)\cdot \varphi = i_X\varphi + \xi\wedge\varphi = 0.
\end{align}
The spinor can in general only be defined locally. This is suitable for our
purposes. More generally, we associate $L$ with a pure spinor line ${\cal U}$ such that 
$\varphi$ is locally a representative of ${\cal U}$ \cite{Gualtieri:2003dx}.  The
spinor satisfies $\GJ_1\cdot\varphi = \i n \varphi$, where $n$ is the complex
dimension of the manifold and the multiplication is given by the action
\begin{align}
  \GJ_1\cdot\varphi = -L\wedge\varphi + \i_P \varphi - J^*\varphi +
  \frac{1}{2}{\rm tr}(J)\varphi.
\end{align}
Here, $J,P,L$ are the components of $\GJ$ as given in \eqref{GJ}. Since $\GJ$ is
integrable, the spinor is pure and satisfies
\begin{align}
  \d \varphi = (u+\xi)\cdot \varphi
\end{align}
for some section $u+\xi$ of $TM\oplus T^*M$.  If $\GJ_1$ was a twisted
generalized complex structure then this equation would have been modified
incorporating $H$:
\begin{align}
  \d_H \varphi = (\d + H\wedge)\varphi = (u+\xi)\cdot \varphi.
\end{align}
Given that $\varphi$ is a pure spinor for the $+\i$ eigenspace of $\GJ_1$, then
\begin{align}
\phi = (1+\tsfrac{1}{2}z\GJ_{3}^{(+)}+\tsfrac{1}{2}w\GJ_{3}^{(-)})\cdot\varphi
\end{align}
is a pure spinor for $\GK$. Since $\GJ_i$ and $\GJt_i$ are integrable by
assumption, $\GK$ is integrable as well. This follows from the fact that the
Nijenhuis concomitants vanish.  Especially, for fix $z,w$, 
\begin{align}
  \d\phi|_{z,w} = (u+\xi)\cdot\phi
\end{align}
for some $u+\xi\in \Gamma(TM\oplus T^*M)$. The bar indicates that the derivative
is taken for fixed values of $z$ and $w$. The generalized complex structure
${\cal J}_S$ is integrable by construction. We can associate to it a pure spinor
$\eta$ such that $(A+b)\cdot \eta = 0$ for sections $A+b$ of $T(S^2\times
S^2)\oplus T^*(S^2\times S^2)$.
Explicitly, $\eta$ is the top-holomorphic form $\eta=\d z \wedge \d w$. Since
$\phi$ is holomorphic in $z,w$, the spinor $\rho =
\phi\wedge \eta$ satisfies 
\begin{align}
  \d(\phi\wedge\eta) &= \d\phi|_{z,w}\wedge\eta + (-1)^{|\phi|}\phi\wedge\d\eta
  + \d z\wedge \nabla_{\partial_z}\phi \wedge \eta
  + \d w\wedge \nabla_{\partial_w}\phi \wedge \eta. \label{dphiwedgeeta}
\end{align}
$\rho$ is a spinor. It is an element of the exterior algebra
$\wedge T^*(M\times S^2\times S^2) = (\wedge T^*M)\wedge (\wedge T^*S^2) \wedge
(\wedge T^*S^2)$. By construction the last two terms in \eqref{dphiwedgeeta}
vanish such that
\begin{align}
  \d \rho &= (X+\xi)\cdot\phi \wedge\eta + 
    (-1)^{|\phi|}\phi\wedge(A+b)\cdot\eta, \cr
  &= (X+\xi+A+b)\cdot\rho.
\end{align}
$\rho$ is a pure spinor for the almost generalized complex structure ${\bf
J}=(\GK({z,w}), {\cal J}_S)$ and we conclude that ${\bf J}$ is integrable.

\section{Discussion} \label{Discussion}
In this short note we showed how generalized hypercomplex geometry emerges as
the target space geometry for the $N=4$ supersymmetric sigma model phase space.
We applied this result to the Hamilton formulation of the $N=(1,1)$
supersymmetric sigma model and combined it with the results of
\cite{Bredthauer:2006hf} to show that the Hamiltonian admits $N=(4,4)$
supersymmetry if the target space is generalized hyperkähler. We defined the
twistor space of generalized complex structures and clarified why the two
parameterizing two-spheres do not parametrize the generalized complex structures
but the supersymmetries in the Lagrangian formulation. Our results fit the
discussion in \cite{Huybrechts:2003ak}. We also discussed the twistor space
construction in terms of pure spinors. This construction should be related to
the deformation complex of generalized complex structures
\cite{Gualtieri:2003dx}. Recently, it has been shown that generalized Kähler
geometry corresponds to a manifest formulation of the $N=(2,2)$ supersymmetric
sigma model \cite{Lindstrom:2005zr}. It would be interesting to relate our
results to the harmonic superspace formulation of $N=(4,4)$ supersymmetry
\cite{Ivanov:2004re} in the same way. Since hyperkähler geometry is always
Calabi-Yau, another interesting open question is how generalized hyperkähler
geometry relates to generalized Calabi-Yau geometry.
\ \\[2ex]

\noindent
{\bf Acknowledgments}\\
The author thanks 
M.~Zabzine and U.~Lindström for many enlightening discussions and comments on
this subject.

%\clearpage
\appendix

\section{Generalized Complex Geometry}
The notation of generalized complex geometry was first introduced by Hitchin
\cite{Hitchin:2004ut} and later clarified by Gualtieri \cite{Gualtieri:2003dx}. 

The vector bundle $T\oplus T^*$ on a complex $d$-dimensional manifold $M$ has a
natural pairing 
\begin{align}
  \langle X+\xi, Y+\eta \rangle = \frac{1}{2}(i_Y \xi + i_X \eta).
\end{align}
The smooth sections of $T\oplus T^*$ have a natural bracket, called the Courant
bracket
\begin{align}
  {[}X+\xi, Y+\eta{]}_c = [X,Y] + L_X\eta - L_Y\xi - \frac{1}{2}\d(i_X \eta -
  i_Y \xi).
\end{align}
It is a natural extension of the Lie-bracket $[\cdot,\cdot]$ on the tangent
bundle onto $T\oplus T^*$. Here, $L_X$ is the Lie derivative with respect to
$X$. This bracket has non-trivial automorphism parametrized by a closed two-form
$b\in\Omega^2_{\rm closed}(M)$ acting on the sections as
\begin{align}
  e^b(X+\xi) = X + (\xi + i_X b).
\end{align}
This transformation is called a $b$-transform and it acts on Courant bracket as
\begin{align}
  {[}e^b(X+\xi),e^b(Y+\eta){]}_c = \e^b{[}X+\xi,Y+\eta{]}_c.
\end{align}
A generalized complex structure is the complex version of two complementary
Dirac structures with $(T\oplus T^*)\otimes \mathbb{C} = L \oplus \bar L$. We can
define it as a map
${\cal J}:(T\oplus T^*)\otimes\mathbb{C} \rightarrow (T\oplus
T^*)\otimes\mathbb{C}$ satisfying
\begin{align}
  {\cal J}^t{\cal I}{\cal J} = {\cal I},&&
  {\cal J}^2=-1_{2d},&&
  \Pi_{\mp}{[}\Pi_{\pm}(X+\xi),\Pi_{\pm}(Y+\eta){]}_c = 0,
\end{align}
where $\Pi_\pm = \frac{1}{2}(1_{2d}\pm {\cal J})$ are projectors on $L$ and
$\bar L$.

\begingroup\raggedright


\begin{thebibliography}{10}

%\cite{Alvarez-Gaume:1980vs}
\bibitem{Alvarez-Gaume:1980vs}
  L.~Alvarez-Gaume and D.~Z.~Freedman,
  ``Ricci Flat Kähler Manifolds And Supersymmetry,''
  Phys.\ Lett.\ B {\bf 94}, 171 (1980).
  %%CITATION = PHLTA,B94,171;%%

%\cite{Bergamin:2004sk}
\bibitem{Bergamin:2004sk}
  L.~Bergamin,
  ``Generalized complex geometry and the Poisson sigma model,''
  Mod.\ Phys.\ Lett.\ A {\bf 20}, 985 (2005)
  [arXiv:hep-th/0409283].
  %%CITATION = HEP-TH 0409283;%%



%\cite{Bredthauer:2005zx}
\bibitem{Bredthauer:2005zx}
  A.~Bredthauer, U.~Lindström and J.~Persson,
  ``First-order supersymmetric sigma models and target space geometry,''
  JHEP {\bf 0601}, 144 (2006)
  [arXiv:hep-th/0508228].
  %%CITATION = HEP-TH 0508228;%%

%\cite{Bredthauer:2006hf}
\bibitem{Bredthauer:2006hf}
  A.~Bredthauer, U.~Lindström, J.~Persson and M.~Zabzine,
  ``Generalized Kähler geometry from supersymmetric sigma models,''
  arXiv:hep-th/0603130.
  %%CITATION = HEP-TH 0603130;%%


%\cite{Buscher:1987uw}
\bibitem{Buscher:1987uw}
  T.~Buscher, U.~Lindström and M.~Ro\v{c}ek,
  ``New supersymmetric sigma models with Wess-Zumino terms,''
  Phys.\ Lett.\ B {\bf 202}, 94 (1988).
  %%CITATION = PHLTA,B202,94;%%

%\cite{Calvo:2005ww}
\bibitem{Calvo:2005ww}
  I.~Calvo,
   ``Supersymmetric WZ-Poisson sigma model and twisted generalized complex
  geometry,''
  Lett.\ Math.\ Phys.\  {\bf 77}, 53 (2006)
  [arXiv:hep-th/0511179].
  %%CITATION = HEP-TH 0511179;%%

%\cite{Davidov:2005}
\bibitem{Davidov:2005}
  J.~Davidov and O.~Mushkarov, 
  ``Twistor spaces of generalized complex structures,''
  arXiv:math.DG/0501396.

%\cite{Davidov:2006}
\bibitem{Davidov:2006}
  J.~Davidov and O.~Mushkarov,    
  ``Twistorial construction of generalized Kähler manifolds,''
   arXiv:math.DG/0607030.

%\cite{Gates:1984nk}
\bibitem{Gates:1984nk}
  S.~J.~Gates, C.~M.~Hull and M.~Ro\v{c}ek,
  ``Twisted Multiplets And New Supersymmetric Nonlinear Sigma Models,''
  Nucl.\ Phys.\ B {\bf 248}, 157 (1984).
  %%CITATION = NUPHA,B248,157;%%

%\cite{Goto:2005}
\bibitem{Goto:2005}
  R.~Goto,
  ``On deformations of generalized Calabi-Yau, hyperKähler, $G_2$ and $Spin(7)$
  structures I,''
  arXiv:math.dg/0512211.
  %%CITATION = MATH-DG 0512211;%%

%\cite{Gualtieri:2003dx}
\bibitem{Gualtieri:2003dx}
  M.~Gualtieri,
  ``Generalized complex geometry,''
  arXiv:math.dg/0401221.
  %%CITATION = MATH-DG 0401221;%%

%\cite{Hitchin:1986ea}
\bibitem{Hitchin:1986ea}
  N.~J.~Hitchin, A.~Karlhede, U.~Lindström and M.~Ro\v{c}ek,
  ``Hyperkähler metrics and supersymmetry,''
  Commun.\ Math.\ Phys.\  {\bf 108} (1987) 535.
  %%CITATION = CMPHA,108,535;%%

%\cite{Hitchin:2004ut}
\bibitem{Hitchin:2004ut}
  N.~Hitchin,
  ``Generalized Calabi-Yau manifolds,''
  Quart.\ J.\ Math.\ Oxford Ser.\  {\bf 54}, 281 (2003)
  [arXiv:math.dg/0209099].
  %%CITATION = MATH-DG 0209099;%%

%\cite{Hitchin:2005cv}
\bibitem{Hitchin:2005cv}
  N.~Hitchin,
  ``Instantons, Poisson structures and generalized Kähler geometry,''
  Commun.\ Math.\ Phys.\  {\bf 265} (2006) 131
  [arXiv:math.dg/0503432].
  %%CITATION = MATH-DG 0503432;%%

%\cite{Howe:1985pm}
\bibitem{Howe:1985pm}
  P.~S.~Howe and G.~Sierra,
  ``Two-Dimensional Supersymmetric Nonlinear Sigma Models With Torsion,''
  Phys.\ Lett.\ B {\bf 148} (1984) 451.
  %%CITATION = PHLTA,B148,451;%%

%\cite{Howe:1988cj}
\bibitem{Howe:1988cj}
  P.~S.~Howe and G.~Papadopoulos,
  ``Further remarks on the geometry of two-dimensional nonlinear sigma models,''
  Class.\ Quant.\ Grav.\  {\bf 5}, 1647 (1988).
  %%CITATION = CQGRD,5,1647;%%

%\cite{Howe:1996kj}
\bibitem{Howe:1996kj}
  P.~S.~Howe and G.~Papadopoulos,
  ``Twistor spaces for HKT manifolds,''
  Phys.\ Lett.\ B {\bf 379} (1996) 80
  [arXiv:hep-th/9602108].
  %%CITATION = HEP-TH 9602108;%%

%\cite{Huybrechts:2003ak}
\bibitem{Huybrechts:2003ak}
  D.~Huybrechts,
  ``Generalized Calabi-Yau structures, K3 surfaces, and B-fields,''
  Int.\ J.\ Math.\  {\bf 16} (2005) 13
  [arXiv:math.ag/0306162].
  %%CITATION = MATH-AG 0306162;%%

%\cite{Ivanov:2001dn}
\bibitem{Ivanov:2001dn}
  S.~Ivanov and I.~Minchev,
   ``Quaternionic Kähler and hyperKähler manifolds with torsion and twistor
  spaces,''
  arXiv:math.dg/0112157.
  %%CITATION = MATH-DG 0112157;%%

%\cite{Ivanov:2004re}
\bibitem{Ivanov:2004re}
  E.~Ivanov and A.~Sutulin,
   ``Diversity of off-shell twisted $(4,4)$ multiplets in $SU(2) \times SU(2)$ harmonic
  superspace,''
  Phys.\ Rev.\ D {\bf 70} (2004) 045022
  [arXiv:hep-th/0403130].
  %%CITATION = HEP-TH 0403130;%%

%\cite{Lindstrom:2004eh}
\bibitem{Lindstrom:2004eh}
  U.~Lindström,
  ``Generalized $N = (2,2)$ supersymmetric non-linear sigma models,''
  Phys.\ Lett.\ B {\bf 587}, 216 (2004)
  [arXiv:hep-th/0401100].
  %%CITATION = HEP-TH 0401100;%%

%\cite{Lindstrom:2004iw}
\bibitem{Lindstrom:2004iw}
  U.~Lindström, R.~Minasian, A.~Tomasiello and M.~Zabzine,
  ``Generalized complex manifolds and supersymmetry,''
  Commun.\ Math.\ Phys.\  {\bf 257}, 235 (2005)
  [arXiv:hep-th/0405085].
  %%CITATION = HEP-TH 0405085;%%

%\cite{Lindstrom:2005zr}
\bibitem{Lindstrom:2005zr}
  U.~Lindstrom, M.~Ro\v{c}ek, R.~von Unge and M.~Zabzine,
  ``Generalized Kähler manifolds and off-shell supersymmetry,''
  arXiv:hep-th/0512164.
  %%CITATION = HEP-TH 0512164;%%

%\cite{Lindstrom:2006ee}
\bibitem{Lindstrom:2006ee}
  U.~Lindström,
   ``A brief review of supersymmetric non-linear sigma models and generalized
  complex geometry,''
  arXiv:hep-th/0603240.
  %%CITATION = HEP-TH 0603240;%%

%\cite{Malikov:2006rm}
\bibitem{Malikov:2006rm}
  F.~Malikov,
  ``Lagrangian approach to sheaves of vertex algebras,''
  arXiv:math.ag/0604093.
  %%CITATION = MATH-AG 0604093;%%

%\cite{Penrose:1976}
\bibitem{Penrose:1976}
  R.~Penrose,
  ``Nonlinear gravitons and curved twistor theory,''
  Gen.\ Rel.\ Grav.\ {\bf 7}, 31 (1976).

%\cite{Salamon:1982}
\bibitem{Salamon:1982}
  S.~Salamon,
  ``Quaternionic Kähler manifolds,''
  Invent.\ Math.\ {\bf 67}, 143 (1982).

%\cite{Salamon:1986}
\bibitem{Salamon:1986}
  S.~Salamon,
  ``Differential geometry of quaternionic manifolds,''
  Ann.\ Sci.\ Ec.\ Norm.\ Sup.\ Paris {\bf 19}, 31 (1986).

%\cite{Zabzine:2005qf}
\bibitem{Zabzine:2005qf}
  M.~Zabzine,
  ``Hamiltonian perspective on generalized complex structure,''
  Commun.\ Math.\ Phys.\  {\bf 263}, 711 (2006)
  [arXiv:hep-th/0502137].
  %%CITATION = HEP-TH 0502137;%%

%\cite{Zabzine:2006uz}
\bibitem{Zabzine:2006uz}
  M.~Zabzine,
  ``Lectures on generalized complex geometry and supersymmetry,''
  arXiv:hep-th/0605148.
  %%CITATION = HEP-TH 0605148;%%

%\cite{Zucchini:2004ta}
\bibitem{Zucchini:2004ta}
  R.~Zucchini,
   ``A sigma model field theoretic realization of Hitchin's generalized  complex
  geometry,''
  JHEP {\bf 0411}, 045 (2004)
  [arXiv:hep-th/0409181].
  %%CITATION = HEP-TH 0409181;%%

%\cite{Zumino:1979et}
\bibitem{Zumino:1979et}
  B.~Zumino,
  ``Supersymmetry And Kähler Manifolds,''
  Phys.\ Lett.\ B {\bf 7}, 203 (1979).
  %%CITATION = PHLTA,B87,203;%%
\end{thebibliography}
\end{document}